\documentclass[preprint,12pt]{elsarticle}

\usepackage{epstopdf}
\usepackage{epsfig}
\usepackage{verbatim}
\usepackage{amssymb} 
\usepackage{amsmath}
\usepackage{amsthm} 

\newcommand {\bw}	{\begin{widetext}}
\newcommand {\ew}	{\end{widetext}}
\newcommand {\be}	{\begin{equation}}
\newcommand {\ee}	{\end{equation}}
\newcommand {\bea}	{\begin{eqnarray}}
\newcommand {\eea}	{\end{eqnarray}}
\newcommand {\mn}[1]	{\langle#1\rangle}

\newcommand {\lbrb}[1]	{\lbrace#1\rbrace}

\newenvironment{color}[3]
{
}{
}

\usepackage{lineno}

\journal{}

\begin{document}

\begin{frontmatter}



\title{Isolation of Flow and Nonflow Correlations by Two- and Four-Particle Cumulant Measurements of Azimuthal Harmonics in $\sqrt{s_{_{\rm NN}}} =$ 200 GeV Au+Au Collisions}




\author{
    N.~M.~Abdelwahab$^{57}$,
    L.~Adamczyk$^{1}$,
    J.~K.~Adkins$^{23}$,
    G.~Agakishiev$^{21}$,
    M.~M.~Aggarwal$^{35}$,
    Z.~Ahammed$^{53}$,
    I.~Alekseev$^{19}$,
    J.~Alford$^{22}$,
    C.~D.~Anson$^{32}$,
    A.~Aparin$^{21}$
% ...
}

\address{$^{1}$AGH University of Science and Technology, Cracow 30-059, Poland}
% ...
\address{$^{19}$Alikhanov Institute for Theoretical and Experimental Physics, Moscow 117218, Russia}
% ...
\address{$^{21}$Joint Institute for Nuclear Research, Dubna, 141 980, Russia}
\address{$^{22}$Kent State University, Kent, Ohio 44242, USA}
\address{$^{23}$University of Kentucky, Lexington, Kentucky, 40506-0055, USA}
% ...
\address{$^{32}$Ohio State University, Columbus, Ohio 43210, USA}
% ...
\address{$^{35}$Panjab University, Chandigarh 160014, India}
% ...
\address{$^{53}$Variable Energy Cyclotron Centre, Kolkata 700064, India}
% ...
\address{$^{57}$World Laboratory for Cosmology and Particle Physics (WLCAPP), Cairo 11571, Egypt}
% ...

\begin{abstract}

A data-driven method was applied to measurements of Au+Au collisions at $\sqrt{s_{_{\rm NN}}} =$ 200 GeV made with the STAR detector at RHIC to isolate pseudorapidity distance $\Delta\eta$-dependent and $\Delta\eta$-independent correlations by using two- and four-particle azimuthal cumulant measurements. We identified a component of the correlation that is $\Delta\eta$-independent, which is likely dominated by anisotropic flow and flow fluctuations. It was also found to be independent of $\eta$ within the measured range of pseudorapidity $|\eta|<1$. The relative flow fluctuation was found to be $34\% \pm 2\% (stat.) \pm 3\% (sys.)$ for particles of transverse momentum $p_{T}$ less than $2$ GeV/$c$. The $\Delta\eta$-dependent part may be attributed to nonflow correlations, and is found to be $5\% \pm 2\% (sys.)$ relative to the flow of the measured second harmonic cumulant at $|\Delta\eta| > 0.7$.   
\end{abstract}


\end{frontmatter}



\section{Introduction}
\label{sec:intro}

Heavy-ion collisions at the  Relativistic Heavy Ion Collider (RHIC) provide a means to study the Quark Gluon Plasma (QGP). In a non-central collision, the overlap region of the colliding nuclei is anisotropic. The energy density gradient converts the initial coordinate-space anisotropy into the final momentum-space anisotropy, generally called anisotropic flow. As the system expands, the coordinate-space anisotropy diminishes. Hence, a measurement of flow is most sensitive to the system at the early stage of the collision \cite{Ollitrault_ecc}. Through measurements of anisotropic flow and comparison with hydrodynamic calculations, properties of the early stage of the collision system may be extracted. One of the important variables, the ratio of the shear viscosity to entropy density of the QGP, was found to be not much larger than the conjectured quantum limit of $1/4 \pi$ \cite{Kovtun}.

The momentum-space anisotropic flow can be characterized by the Fourier coefficients, $v_{n}$, of the outgoing particle azimuthal ($\phi$) distribution \cite{Voloshin_Zhang}:  
\bea \label{eq:Fourier}
\displaystyle \dfrac{dN}{d\phi}  \propto   1+ \sum_{n = 1}^{\infty} 2 v_{n} \cos n ( \phi-\psi_{n}),
\eea
where the participant plane is characterized by the angle $\psi_{n}$, given by the initial participant nucleon (or parton) configuration \cite{PHOBOS_v2}. The higher harmonics can arise from initial fluctuations such that $\psi_{n}$ is not necessarily the same for different $n$. Because $\psi_{n}$ is not experimentally accessible, the event plane, constructed from final particle momenta, is used as a proxy for the intial state participant plane. The determination of the anisotropic flow uses particle correlations that are, however, contaminated by intrinsic particle correlations unrelated to the participant plane. Those correlations are generally called nonflow and are due to jet fragmentation and final state interactions, such as quantum statistics, Coulomb and strong interactions, and resonance decays \cite{Bor}.

Similarly, two- and multi-particle correlations are also used to measure anisotropy \cite{Wang_2part, Bor}. For example, the two-particle correlation is given by:
\bea \label{eq:2pc}
\displaystyle \dfrac{d N}{d \Delta\phi } \propto 1 + \sum_{n=1}^{\infty} 2 V_{n}\lbrace2\rbrace \cos n \Delta \phi,
\eea
where $\Delta\phi$ is the azimuthal angle between the two particles. In the absence of nonflow, Eq.~\eqref{eq:2pc} follows from Eq.~\eqref{eq:Fourier} with $V_{n}\lbrb{2} = v_{n,\alpha} v_{n, \beta}$ (where $\alpha$, $\beta$ stand for the two particles). Otherwise, $V_{n}\lbrb{2} = v_{n, \alpha} v_{n, \beta} + \delta_{n}$, where $\delta_{n}$ is the nonflow contribution. Since even a small uncertainty in flow can introduce a large error in the extracted shear viscosity \cite{song}, it is important to separate nonflow contributions from flow measurements.


This article describes a method used to separate flow and nonflow in a data-driven way, with minimal reliance on models. We measure two- and four-particle cumulants with different pseudorapidity ($\eta$) combinations. By exploiting the symmetry of the average flow in $\eta$ at midrapidity in symmetric heavy-ion collisions, we separate $\Delta\eta$-independent and $\Delta\eta$-dependent contributions. We associate the $\Delta\eta$-independent part with flow, while the $\Delta\eta$-dependent part is associated with nonflow. This is because flow is an event-wise many-particle azimuthal correlation, reflecting properties on the single-particle level. By contrast, nonflow is a few-particle azimuthal correlation that depends on the $\Delta\eta$ distance between the particles.

This article is organized as follows: Section~\ref{sec:exp} gives the experimental details and the criteria for the data selection. Section~\ref{sec:method} gives two- and four-particle cumulant results and the separation of $\Delta\eta$-independent and $\Delta\eta$-dependent components. Section~\ref{sec:result} associates the $\Delta\eta$-independent part with flow and the $\Delta\eta$-dependent part with nonflow, and further discusses flow, flow fluctuation and nonflow.

\section{Data Analysis}
\label{sec:analysis}

\subsection{Experiment Details and Data Selection}
\label{sec:exp}

This analysis principally relies on the STAR Time Projection Chamber (TPC) \cite{STAR_TPC}. A total of 25 million Au+Au collisions at $\sqrt{s_{_{\rm NN}}} =$ 200 GeV, collected with a minimum bias trigger in 2004, were used. The events selected were required to have a primary event vertex within $|z_{vtx}|< 30$ cm along the beam axis ($z$) to ensure nearly uniform detector acceptance. The centrality definition was based on the raw charged particle multiplicity within $|\eta|<0.5$ in TPC.  The charged particle tracks used in the analysis were required to satisfy the following conditions: the transverse momentum $0.15 < p_{T} < 2 $ GeV$/c$ to remove high $p_{T}$ particles from the jets; the distance of closest approach to the event vertex $|dca| < 3$ cm to ensure that the particles are from the primary collision vertex instead of a secondary particle decay vertex; the number of fit points along the track greater than 20, and the ratio of the number of fit points along the track to the maximum number of possible fit points larger than 0.51 for good primary track reconstruction \cite{Levente}. For the particles used in this paper, the pseudorapidity region was restricted to $|\eta|<1$.



\subsection{Analysis Method}
\label{sec:method}

In this analysis, the anisotropy was calculated by the two- and four-particle Q-cumulant method using unit weight with a non-uniform acceptance correction \cite{Bilandzic_directcalc}. By using the moment of the flow vector, this method makes multi-particle cumulant calculation faster without going over pair or a higher multiplet loop. The non-uniform acceptance correction for 20-30\% centrality was 0.7\% for the second harmonic two-particle cumulant $V_{2}\lbrb{2}$, and 0.5\% for the square root of the second harmonic four-particle cumulant $V_{2}^{1/2}\lbrb{4}$. The largest acceptance correction was 1.8\% for $V_{2}\lbrb{2}$ at the most central, and 1\% for $V_{2}^{1/2}\lbrb{4}$ at the most peripheral collisions.

The two-particle cumulant, with one particle at pseudorapidity $\eta_{\alpha}$ and another at $\eta_{\beta}$, is \cite{Art}
\bea	
V \lbrace 2 \rbrace  & \equiv & \langle\langle e^{i (\phi_{\alpha}-\phi_{\beta})} \rangle\rangle = \mn{v(\eta_{\alpha}) v(\eta_{\beta})} + \delta(\Delta\eta) \nonumber \\
	  & \equiv & \mn{v (\eta_{\alpha})} \mn{v (\eta_{\beta})} + \sigma (\eta_{\alpha}) \sigma (\eta_{\beta}) + \sigma ' (\Delta\eta) + \delta (\Delta \eta), \nonumber \\  \label{EqV2}
\eea
where $\Delta\eta=|\eta_{\beta}-\eta_{\alpha}|$. The double brackets represent the average over particle pairs and the average over events, while the single brackets are for the average over events only. The harmonic number $n$ is suppressed to lighten the notation. The average flow, $\mn{v}$, which is the anisotropy parameter with respect to the participant plane, and the flow fluctuation, $\sigma$, are only functions of $\eta$, because flow reflects the property on the single-particle level. Both $\mn{v}$ and $\sigma$ are $\Delta\eta$-independent quantities. However, because of the way the two-particle cumulant is measured, i.e. by two-particle correlation, there could exist a $\Delta\eta$-dependent flow fluctuation component. For example, the event planes determined by particles at different $\eta$'s can be different \cite{eta-dep}. In Eq.~\eqref{EqV2}, $\sigma'$ denotes this $\Delta\eta$-dependent part of the flow fluctuation. The $\delta$ is the contribution from nonflow, which is generally a function of $\Delta\eta$, but may also depend on $\eta$. 

For the four-particle cumulant, we take two particles at $\eta_{\alpha}$ and another two at $\eta_{\beta}$. For easier discussion, we take the square root of the four-particle cumulant, which has the same order in $\mn{v}$ as the two-particle cumulant. It is given by
\bea  
V^{\frac{1}{2}} \lbrace 4 \rbrace  & \equiv & \sqrt{\mn{\mn{e^{i (\phi_{\alpha}+\phi_{\alpha}-\phi_{\beta}-\phi_{\beta})} }} } \nonumber \\ 
& \approx & \mn{v (\eta_{\alpha})} \mn{v (\eta_{\beta})} - \sigma (\eta_{\alpha}) \sigma (\eta_{\beta}) - \sigma ' (\Delta\eta),\nonumber \\
\label{EqV4}
\eea
where the approximation is that the flow fluctuation is relatively small compared with the average flow \cite{Ollitrault_fluc_nonflow}. In $V^{1/2}\lbrb{4}$, the contribution from the two-particle correlations due to nonflow effects is suppresed, while the contribution from the four-particle correlations due to nonflow effects $\propto 1/{M^{3}}$ ($M$ is multiplicity) and is, therefore, negligible \cite{STAR_v2centrality,STAR_v2}. The fluctuation gives negative contribution to $V^{1/2}\lbrb{4}$, while positive to $V\lbrb{2}$.

\begin{figure*}[htb]   
 \centering
 \includegraphics[width=0.3\textwidth]{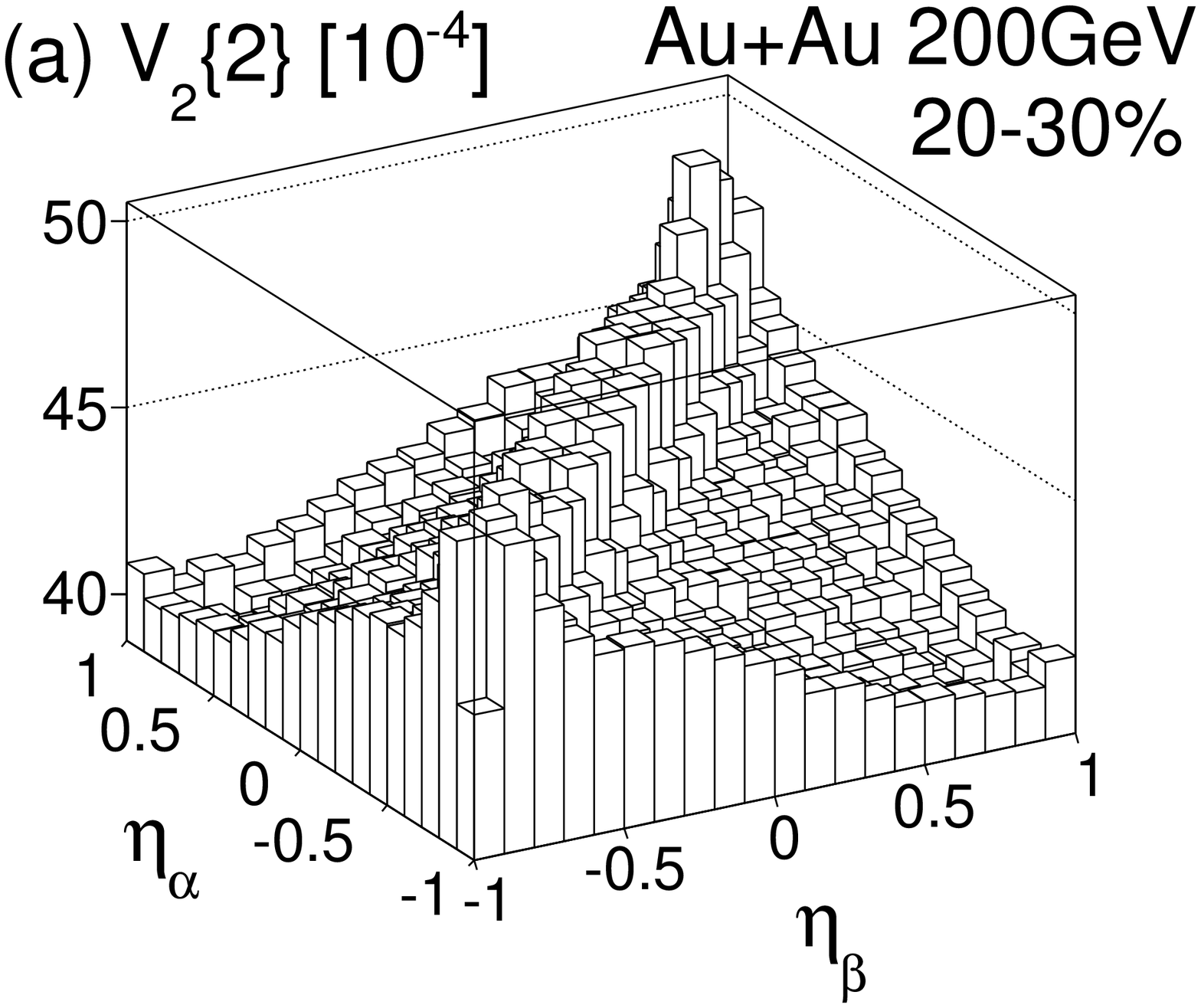} 
     \includegraphics[width=0.3\textwidth]{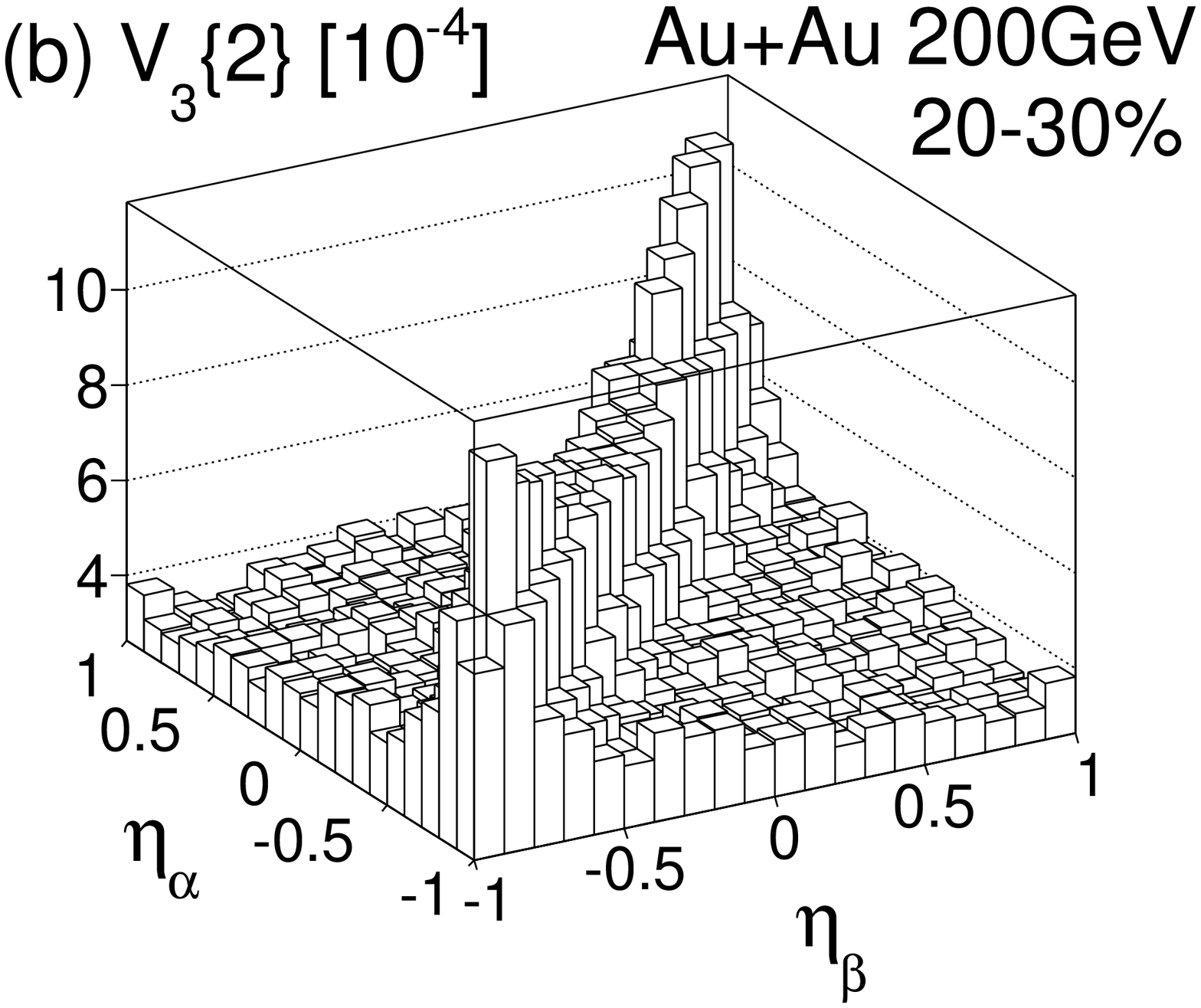} 
     \includegraphics[width=0.3\textwidth]{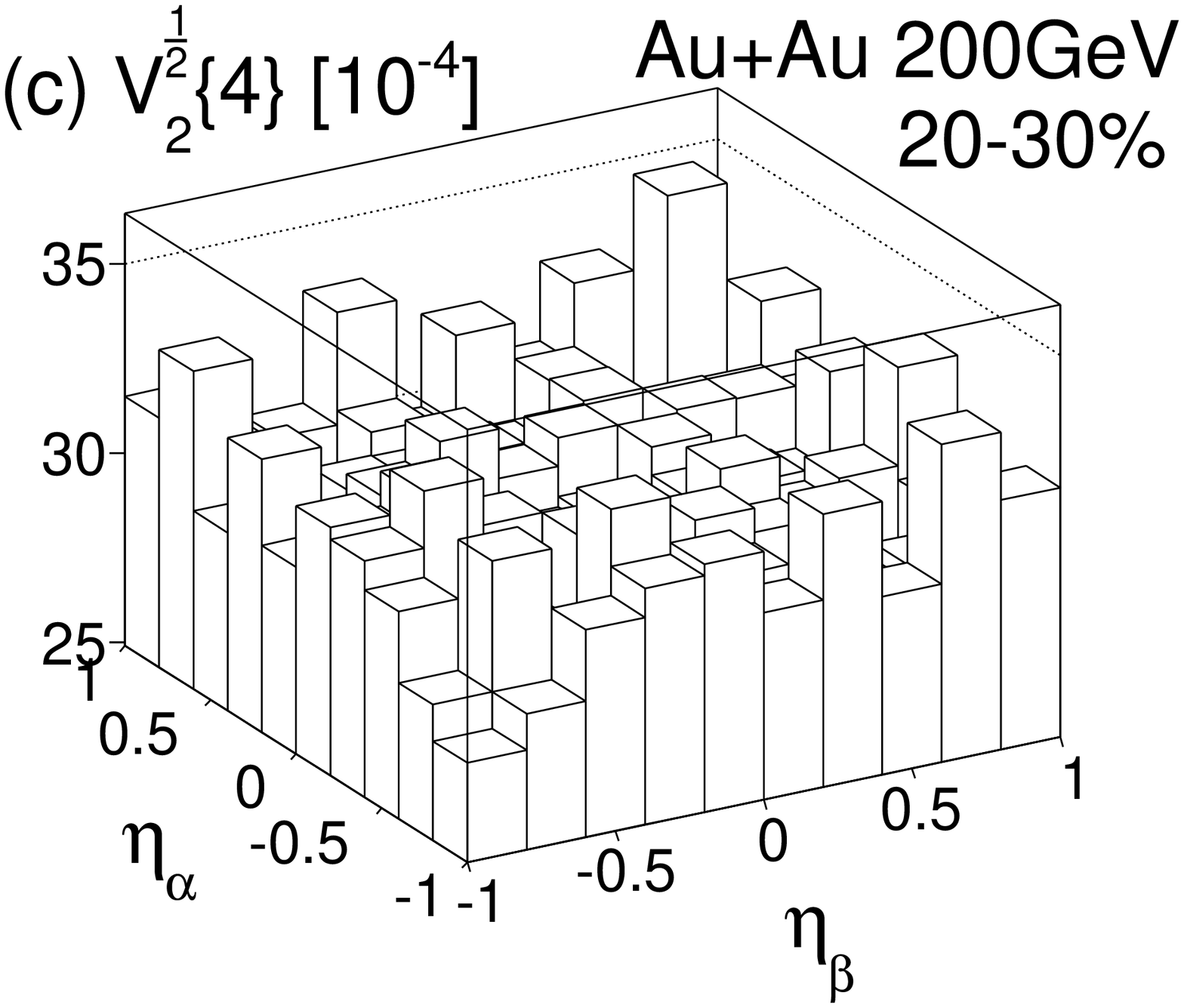} 
  \caption{The second (a) and third (b) harmonic two-particle cumulants for $(\eta_{\alpha},\eta_{\beta})$ pairs and the second harmonic four-particle cumulant for $(\eta_{\alpha},\eta_{\alpha},\eta_{\beta},\eta_{\beta})$ quadruplets for 20-30\% central Au+Au collisions at $\sqrt{s_{_{\rm NN}}} =$ 200 GeV.}
\label{figlegoV2V4}
\end{figure*}

The two- and four-particle cumulants were measured for various $(\eta_{\alpha},\eta_{\beta})$ pairs and quadruplets. Figure \ref{figlegoV2V4} shows the results for 20-30\% central Au+Au collisions. Panels (a) and (b) are the two-particle second  and third harmonic cumulants, $V_{2}\lbrb{2}(\eta_{\alpha},\eta_{\beta})$ and $V_{3}\lbrb{2}(\eta_{\alpha},\eta_{\beta})$, respectively. Panel (c) is the square root of the four-particle second harmonic cumulant, $V^{1/2}_{2}\lbrb{4}(\eta_{\alpha},\eta_{\alpha},\eta_{\beta},\eta_{\beta})$. We observe from Fig.~\ref{figlegoV2V4} that $V_{2}\lbrb{2}$ decreases as the gap between $\eta_{\alpha}$ and $\eta_{\beta}$ increases. Since the track merging affects the region $|\Delta\eta|<0.05$, the $V_{n}\lbrb{2}$ and $V_{n}\lbrb{4}$ points along the diagonal were excluded from further analysis. $V_{3}\lbrb{2}$ follows the same trend, but the magnitude is smaller. $V_{3}\lbrb{2}$ decreases more rapidly with $\Delta\eta$ than does $V_{2}\lbrb{2}$. $V_{2}^{1/2}\lbrb{4}$ is roughly constant and the magnitude is smaller than that of $V_{2}\lbrb{2}$ which is consistent with our expectation that $V_{2}^{1/2}\lbrb{4}$ is less affected by the nonflow and the flow fluctuation is negative in $V_{2}^{1/2}\lbrb{4}$.

In order to extract the values of the average flow, $\mn{v}$, the $\Delta\eta$-dependent and $\Delta\eta$-independent flow fluctuations, $\sigma'$ and $\sigma$,  and the nonflow contribution, $\delta$, we follow an analysis method described in  Ref.~\cite{Xu}. By taking the difference between cumulants $V\lbrace 2 \rbrace$ at $(\eta_{\alpha}, \eta_{\beta})$ and $(\eta_{\alpha}, -\eta_{\beta})$, we have,
\bea   
\Delta V \lbrace 2 \rbrace & \equiv & V \lbrace 2 \rbrace (\eta_{\alpha},\eta_{\beta})- V \lbrace 2 \rbrace (\eta_{\alpha},-\eta_{\beta}) \nonumber \\
 & \equiv & V \lbrace 2 \rbrace (\Delta\eta_{1})- V \lbrace 2 \rbrace (\Delta\eta_{2}) = \Delta \sigma '  + \Delta \delta, \label{EqDV2} 
\eea
where $\eta_{\alpha}<\eta_{\beta}<0$ or $0<\eta_{\beta}<\eta_{\alpha}$ is required. Similarly, this difference for $V^{1/2} \lbrace 4 \rbrace$ yields,
\bea
\Delta V^{\frac{1}{2}} \lbrace 4 \rbrace & \equiv & V^{\frac{1}{2}} \lbrace 4 \rbrace (\eta_{\alpha},\eta_{\beta})- V^{\frac{1}{2}} \lbrace 4 \rbrace (\eta_{\alpha},-\eta_{\beta})  \nonumber \\
 & \equiv & V^{\frac{1}{2}} \lbrace 4 \rbrace (\Delta\eta_{1})- V^{\frac{1}{2}} \lbrace 4 \rbrace (\Delta\eta_{2}) \approx -\Delta \sigma '. \label{EqDV4}
\eea
Here $
\Delta\eta_{1}  \equiv \eta_{\beta} - \eta_{\alpha}  , \ 
\Delta\eta_{2} \equiv -\eta_{\beta} - \eta_{\alpha} \label{EqDh}, \ 
\Delta\sigma'=\sigma'(\Delta\eta_{1})-\sigma'(\Delta\eta_{2})$, and $\Delta\delta = \delta(\Delta\eta_{1})-\delta(\Delta\eta_{2}) . \ 
$
In symmetric heavy-ion collisions, the difference of the two $\Delta\eta$-independent terms in Eq.~\eqref{EqV2} and ~\eqref{EqV4} is zero. Therefore the differences in Eqs.~\eqref{EqDV2} and \eqref{EqDV4} depend only on the $\Delta\eta$-dependent terms: flow fluctuation $\Delta \sigma'$ and nonflow $\Delta \delta$.  

Our goal is to parameterize the flow fluctuation $\Delta \sigma'$ and nonflow $\Delta \delta$. The following part of this section is organized in this way: First, we discuss the empirical functional form for
\bea
\displaystyle D(\Delta\eta)&=&\sigma'(\Delta\eta)+\delta(\Delta\eta), \label{Eqd_2}
\eea
obtained from $\Delta V_{2}\lbrb{2}$ data. Second, we give the $\sigma'$ result from $\Delta V_{2}^{1/2}\lbrb{4}$ . Using $D$ and $\sigma'$, $\delta$ can be determined. Third, we discuss how to obtain $\mn{v}$ and $\sigma$. 


\begin{figure*}[htb]	
 \centering
  \includegraphics[width=0.3\textwidth]{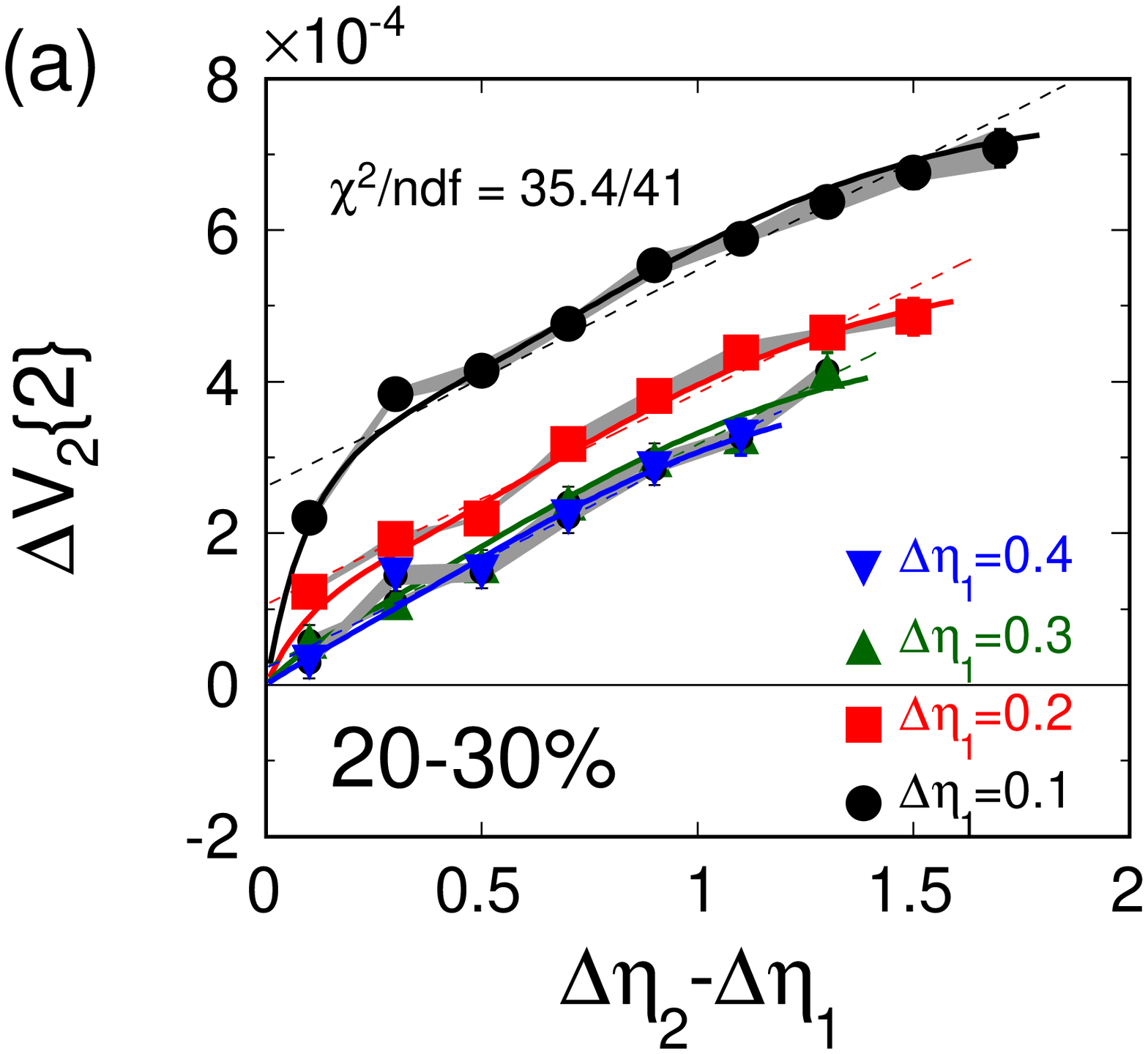} 
  \includegraphics[width=0.3\textwidth]{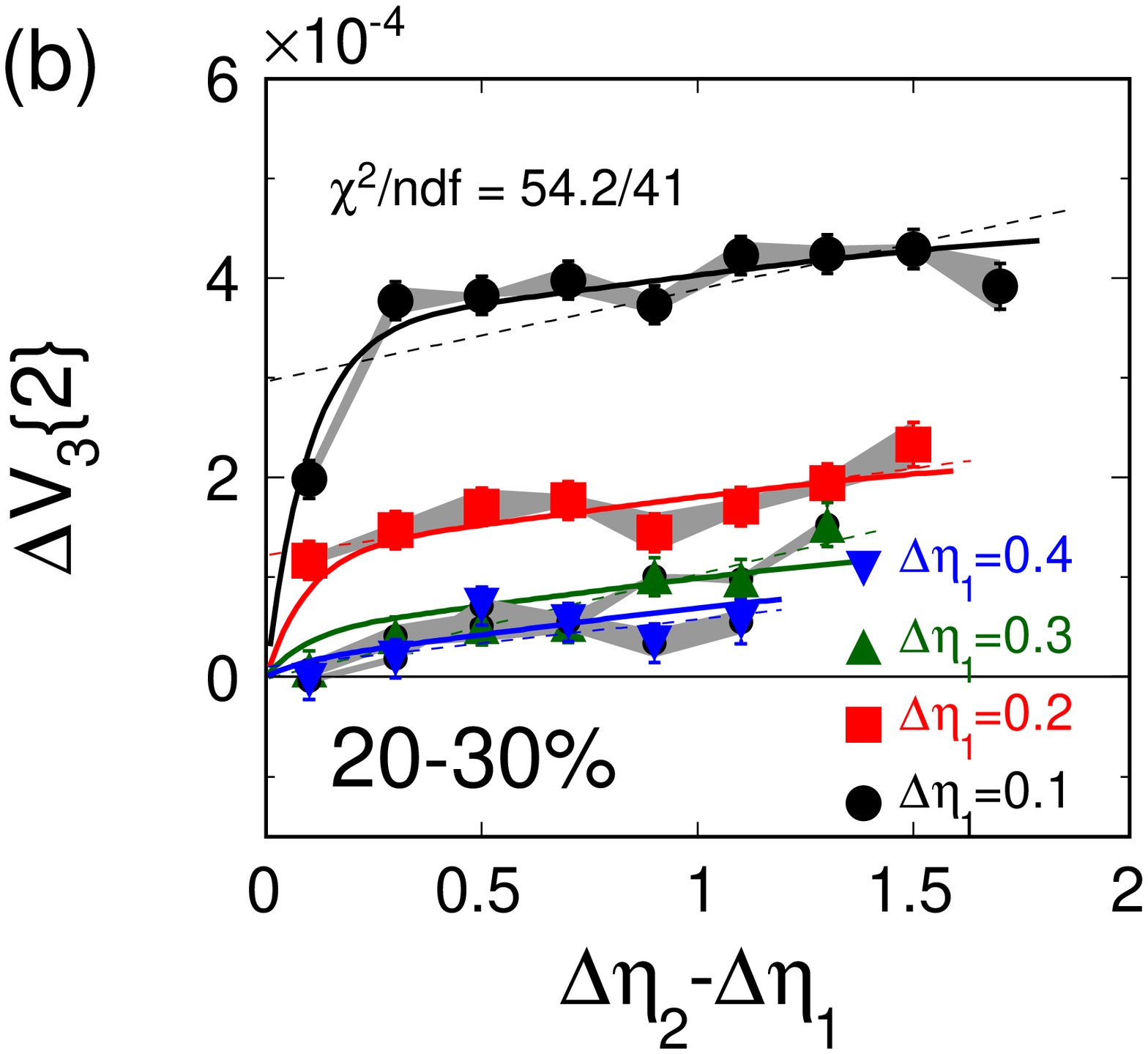} 
  \includegraphics[width=0.3\textwidth]{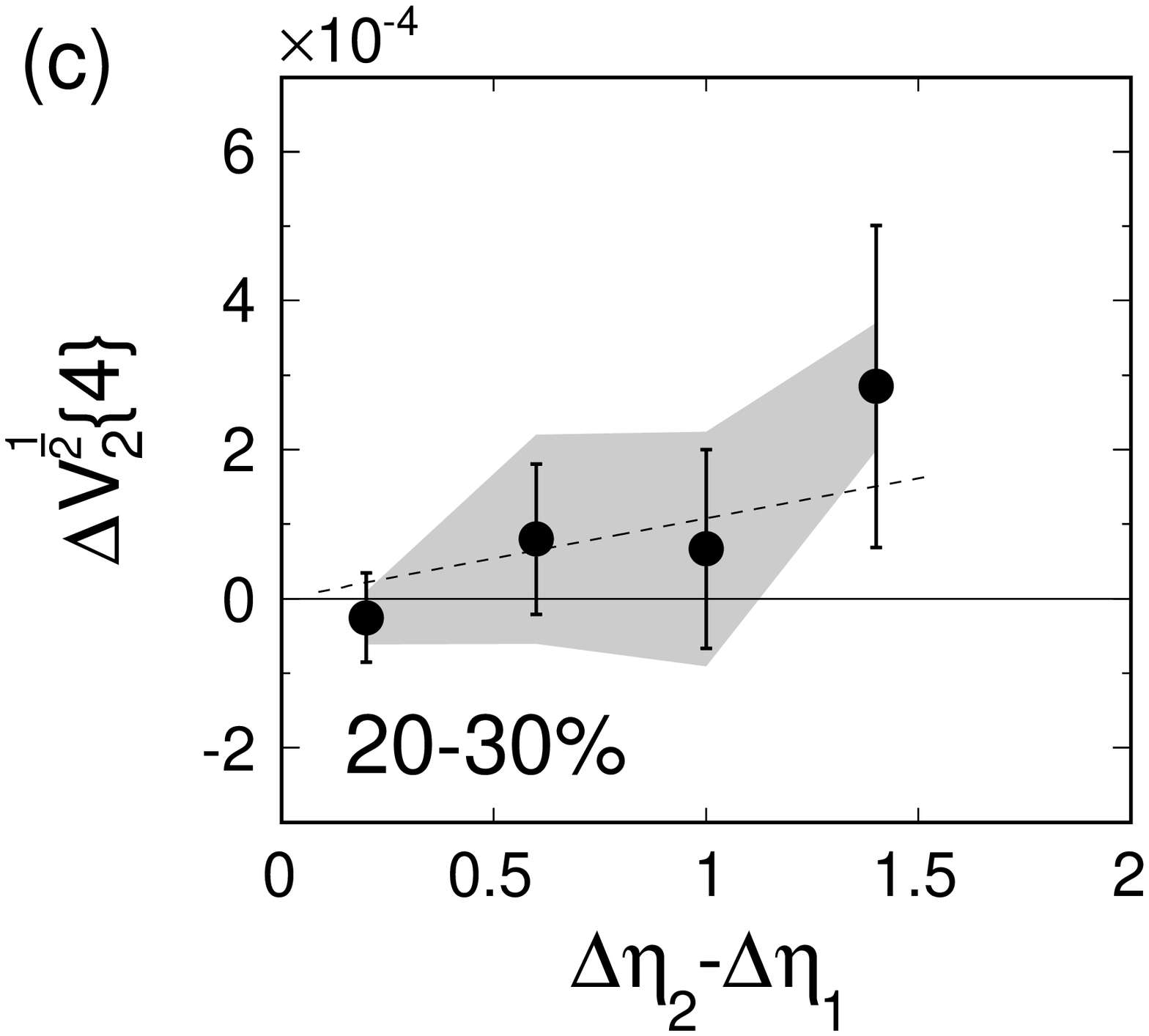} 
  \caption{The (a) $V_{2}\lbrb{2}$ and (b) $V_{3}\lbrb{2}$ difference between the pairs at $(\eta_{\alpha}, \eta_{\beta})$ and $(\eta_{\alpha}, -\eta_{\beta})$. The dashed lines are linear fits for each data set of $\Delta\eta_{1}$ value separately. The solid curves are a single fit of Eq.~\eqref{Eqd} to all data points with different $\Delta\eta_{1}$. (c) The $V_{2}^{1/2}\lbrb{4}$ difference between quadruplets at $(\eta_{\alpha},\eta_{\alpha}, \eta_{\beta}, \eta_{\beta})$ and $(\eta_{\alpha},\eta_{\alpha}, -\eta_{\beta}, -\eta_{\beta})$. The dashed line is a linear fit to the data points. The gray band is the systematic error. The data are from 20-30\% central Au+Au collisions at $\sqrt{s_{_{\rm NN}}} =$ 200 GeV.}
  \label{figfit}
\end{figure*}

The behavior of $\Delta V_{2}\lbrb{2}$ data suggests that $D$ can be parameterized as 
\bea
\displaystyle D(\Delta\eta) &=& a \exp\left({-\frac{\Delta\eta}{b}}\right) + A \exp\left({-\frac{\Delta\eta^{2}}{2 \sigma^{2}}}\right), \label{Eqd} 
\eea
so that
\bea
\Delta V\lbrb{2}&=&D(\Delta\eta_{1})-D(\Delta\eta_{2}) \nonumber\\
&=&\left[a\exp\left({\frac{-\Delta\eta_{1}}{b}}\right)+A\exp\left({\frac{-\Delta\eta_{1}^{2}}{2\sigma^{2}}}\right)\right] \nonumber \\
 &-& \left[a \exp\left({\frac{-\Delta\eta_{2}}{b}}\right) + A \exp\left({\frac{-\Delta\eta_{2}^{2}}{2 \sigma^{2}}}\right) \right], \nonumber\\
\label{EqDd}
\eea
follows from Eq.~\eqref{EqDV2}. Here is how this functional form is chosen. The measured two-particle second harmonic cumulant difference $\Delta V_{2} \lbrb{2}$ is shown in Fig.~\ref{figfit}(a). The data for each $\Delta\eta_{1}$ value appears to be linear in $\Delta\eta_{2}-\Delta\eta_{1}$ except near $\Delta\eta_{1} = \Delta\eta_{2}$ as shown by dashed lines in Fig.~\ref{figfit} (a) and (b). Moreover, the magnitude of $\Delta V_{2}\{2\}$ decreases with increasing $\Delta\eta_{1}$. Linear fits indicate that the intercept decreases exponentially with increasing $\Delta\eta_{1}$, and the slopes are all similar. So we can describe this behavior mathematically as $a \exp(-\frac{\Delta\eta_{1}}{b})$$+k (\Delta\eta_{2}$$-\Delta\eta_{1})$. In order to express the measured two-particle cumulant difference in the form of $D(\Delta\eta_{1})-D(\Delta\eta_{2})$ $= (\sigma'(\Delta\eta_{1})$$+\delta(\Delta\eta_{1}))$$- (\sigma'(\Delta\eta_{2})$$+\delta(\Delta\eta_{2}))$ $=$$(\sigma'(\Delta\eta_{1})$$-\sigma'(\Delta\eta_{2}))$$+(\delta(\Delta\eta_{1})$$-\delta(\Delta\eta_{2}))$, we make two improvements to our initial guess of the $D(\Delta\eta)$ function. First, we add a term $a \exp(-\frac{\Delta\eta_{2}}{b})$ that is small for all data with $\Delta\eta_{2}$ significantly larger than $\Delta\eta_{1}$. Second, because the linear term is unbounded in $\Delta\eta_{1}$ and $\Delta\eta_{2}$, we choose to replace it with the subtraction of two wide Gaussian terms. The Gaussian functions tend to zero as the exponents become large, consistent with the behavior of nonflow. The measured two-particle cumulant difference can then be described by Eq.~\eqref{EqDd}.
There are four parameters in Eq.~\eqref{EqDd} $a, A, b$, and $\sigma$, that were determined by fitting Eq.~\eqref{EqDd} simultaneously to all measured two-particle cumulant difference data points of different $\Delta\eta_{1}$. The fit results are shown in Fig.~\ref{figfit}(a) as the solid curves with $\chi^2/$ndf $\approx 1$. The parameterization is valid within the fitting errors. The same procedure was repeated for the third harmonic $V_{3}\lbrb{2}$ as shown in Fig.~\ref{figfit}(b). The fit results give the $\Delta\eta$-dependent part of the two-particle cumulant as Eq.~\eqref{Eqd}. Thus, the form of the function $D$ is data-driven.

We then follow a similar procedure on the measured difference of the square root of the four-particle cumulant, Eq.~\eqref{EqV4}. We fit the $\Delta V^{1/2}_{2}\lbrb{4}$ $= \sigma'(\Delta\eta_{1})-\sigma'(\Delta\eta_{2})$ by a linear function $k'(\Delta\eta_{2}-\Delta\eta_{1})$, as shown in Fig.~\ref{figfit}(c). The slope $k'$ from the fit is $(1.1\pm0.8)\times10^{-4}$. In Fig.~\ref{figfit}(c), each data point is the average of $\Delta V_{2}^{1/2}\lbrb{4}$ for all $\Delta \eta_{1}$ at same $\Delta\eta_{2}-\Delta\eta_{1}$ value. With the $\sigma'(\Delta\eta)$ result, the contribution from nonflow, $\delta$, can then also be determined from Eq.~\eqref{Eqd_2}.

Subtracting the parameterized $D$ of Eq.~\eqref{Eqd} from the measured two-particle cumulants, $V_{2}\lbrb{2}$ and $V_{3}\lbrb{2}$, yields, from Eq.~\eqref{EqV2}, the $\Delta\eta$-independent terms $\mn{v^{2}} \equiv \mn{v}^{2} + \sigma^{2}$. Employing also $V^{1/2}\lbrb{4}$ from Eq.~\eqref{EqV4}, the values of $\mn{v}$ and $\sigma$ may be individually determined. 

\begin{figure*}[htb]	
 \centering
 \begin{tabular}{cc}
\includegraphics[width=0.4\textwidth]{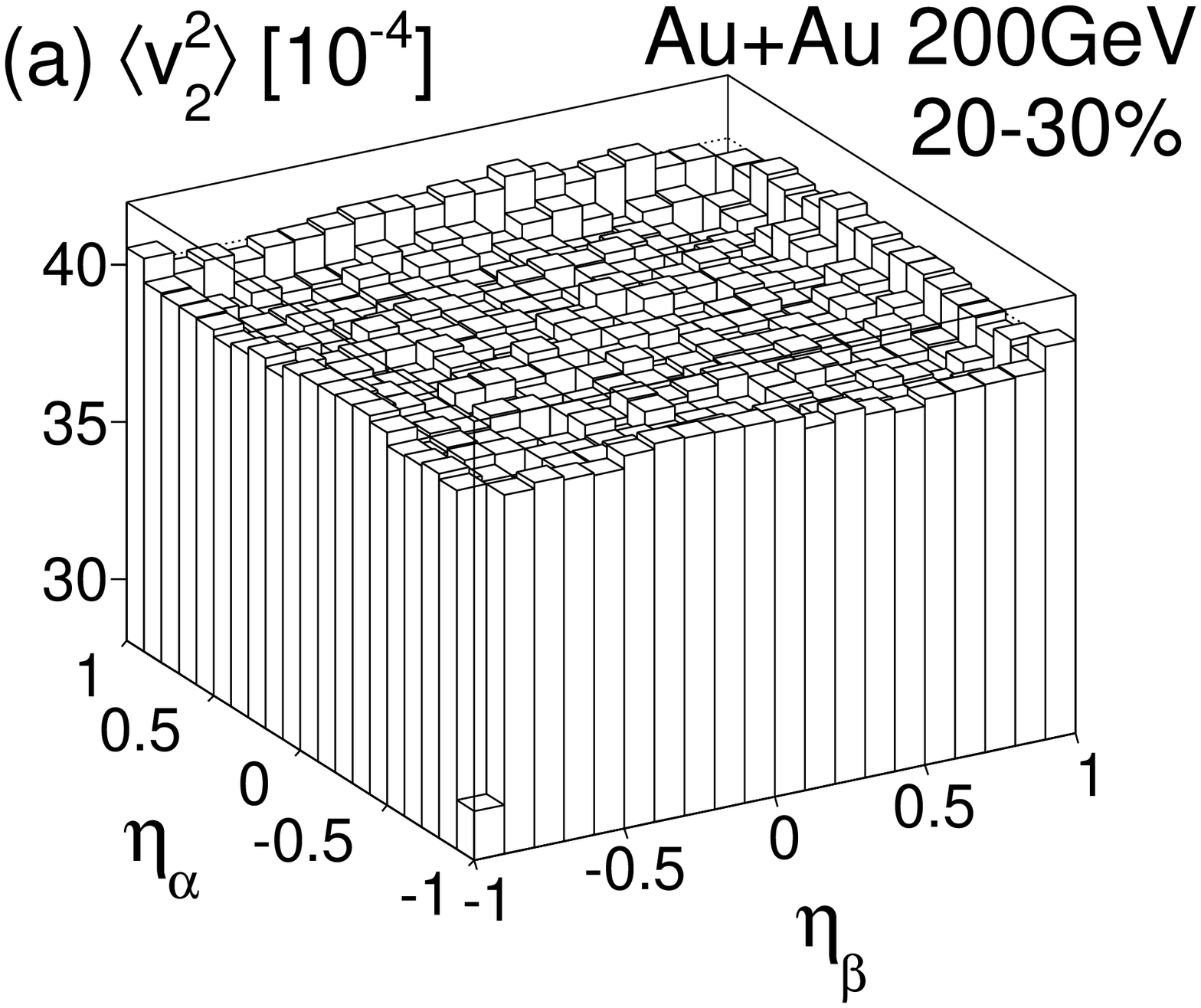} & \includegraphics[width=0.4\textwidth]{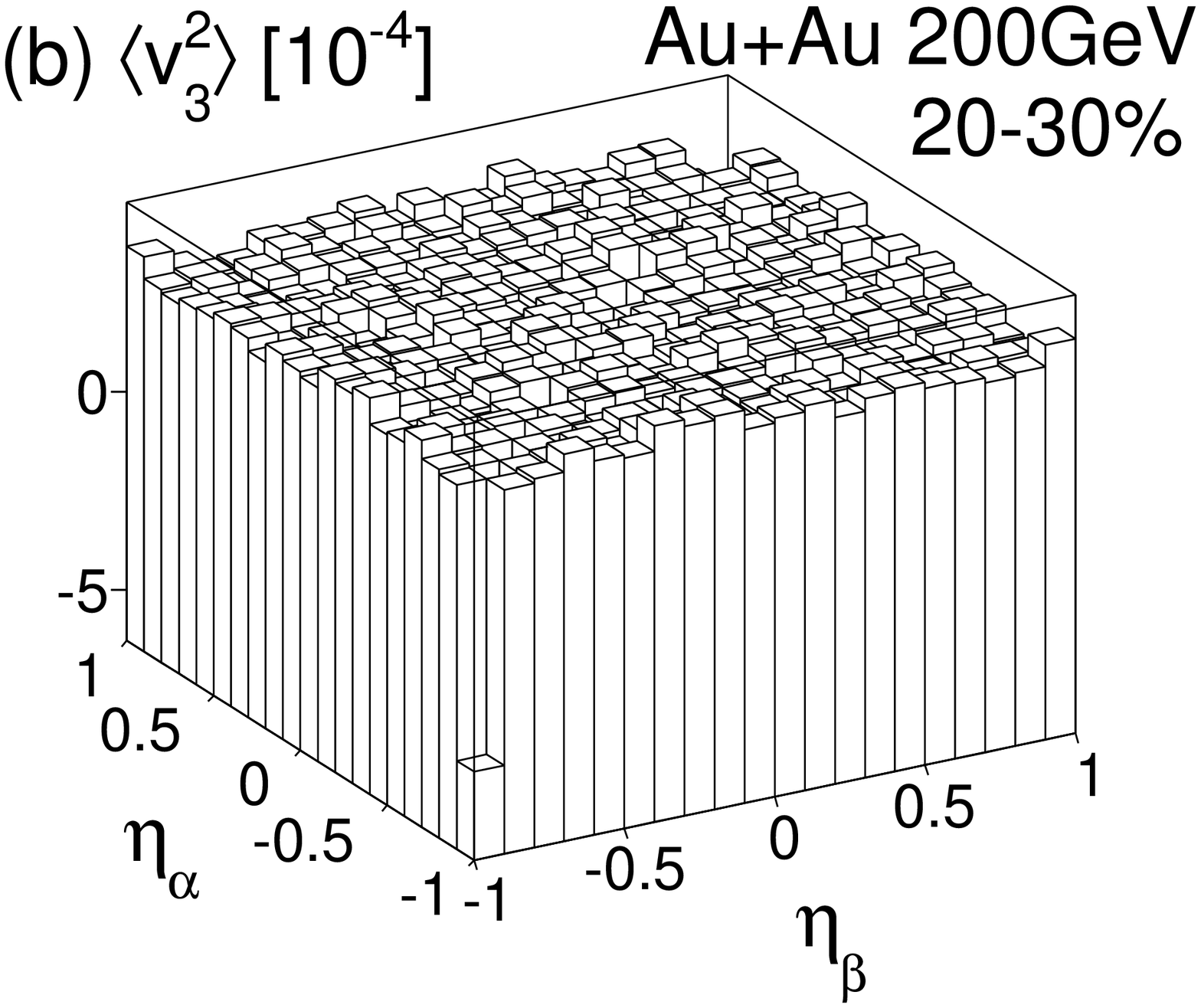} \\
\includegraphics[width=0.4\textwidth]{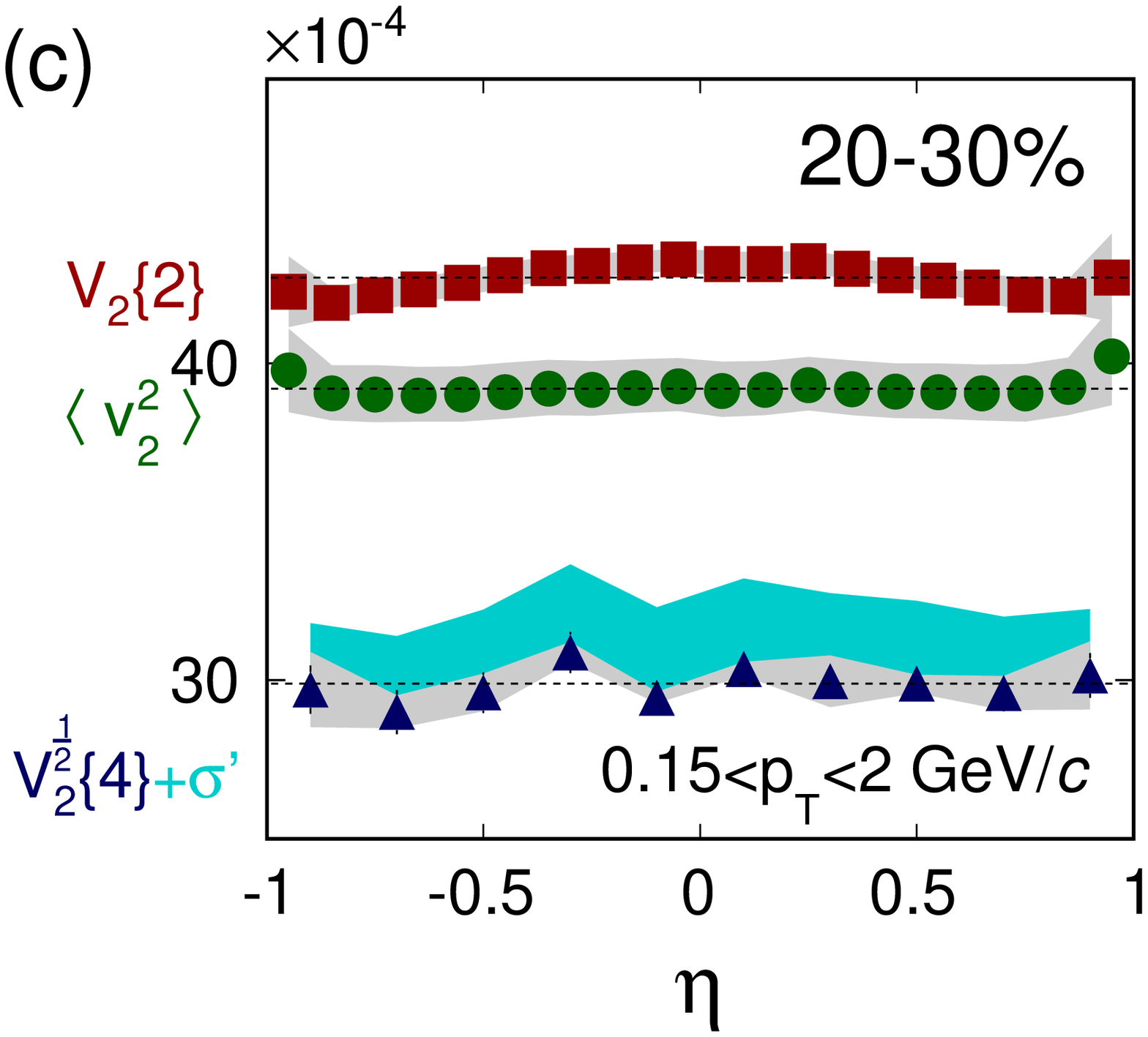} & \includegraphics[width=0.4\textwidth]{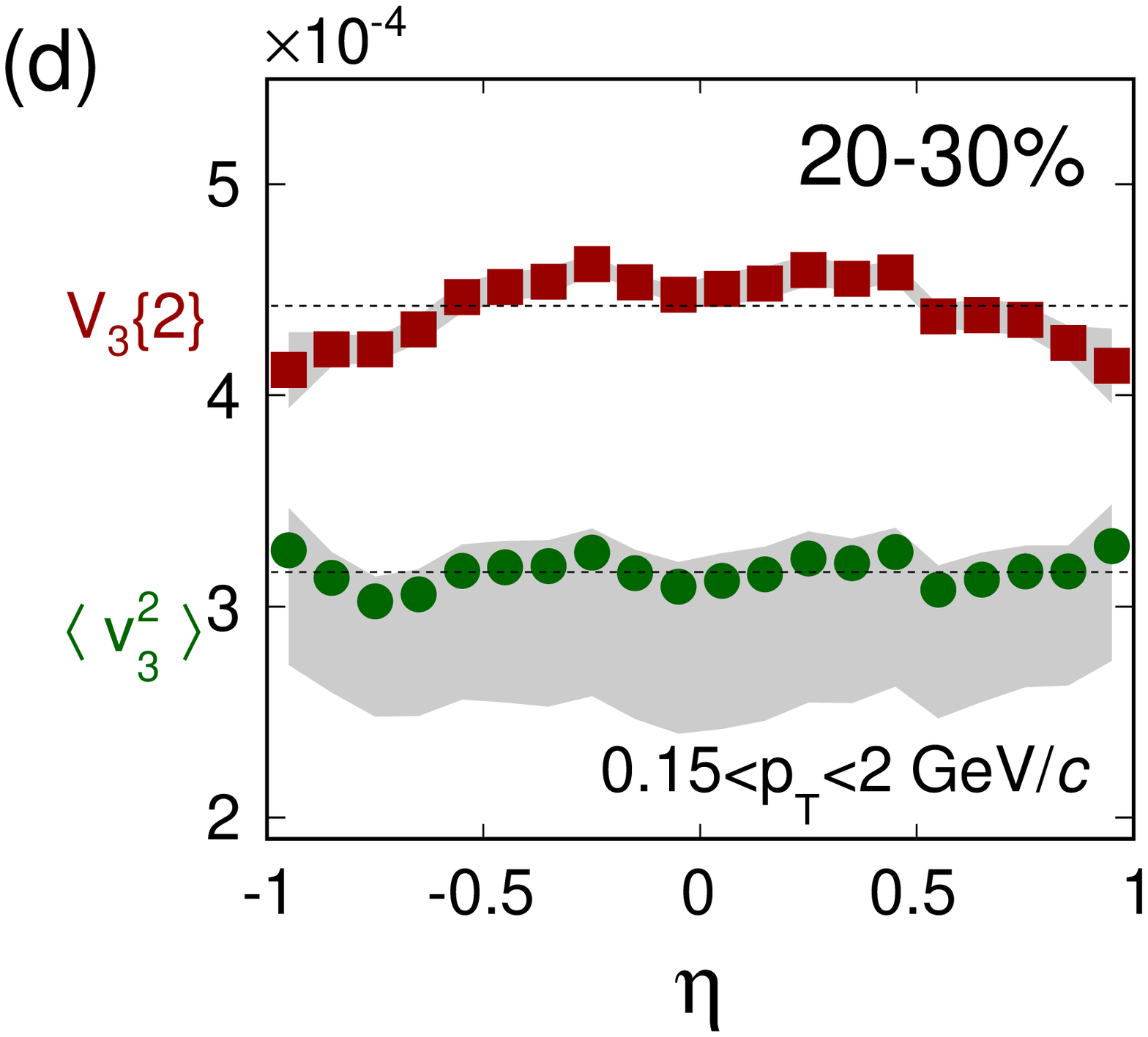} \\
\end{tabular}
  \caption{The decomposed $\mn{v^{2}}=\mn{v}^{2} + \sigma^{2}$ for the second (a) and third (b) harmonics for $(\eta_{\alpha},\eta_{\beta})$ pairs. (c): The two- and four-particle cumulants, $V_{2}\lbrb{2}$ (solid red squares) and $V_{2}^{1/2}\lbrb{4}$ (solid blue triangles), and the decomposed $\mn{v^{2}_{2}}$ (solid green dots) as a function of $\eta$ for one particle while averaged over $\eta$ of the partner particle. The cyan band on top of $V_{2}^{1/2}\lbrb{4}$ points present $V_{2}^{1/2}\lbrb{4}+\sigma'$. (d): $V_3\lbrb{2}$ (solid red squares) and $\mn{v^{2}_{3}}$ (solid green dots) as a function of $\eta$. The dashed lines are the mean value averaged over $\eta$ for 20-30\% central Au+Au collisions at $\sqrt{s_{_{\rm NN}}} =$ 200 GeV.}
\label{figv}
\end{figure*}

\subsection{Systematic Uncertainties}	
\label{sec:syserr}
   The systematic errors for $V\lbrb{2}$ and $V^{1/2}\lbrb{4}$ are estimated by varying event and track quality cuts: the primary event vertex to $|z_{vtx}|<25$ cm; the number of fit points along the track greater than 15; the distance of closest approach to the event vertex $|dca|<2$ cm. The systematic errors for events at 20-30\% centrality were found to be 1\% for $V_{2}\lbrb{2}$ and 2\% for $V_{2}^{1/2}\lbrb{4}$, and the same order of magnitude for other centralities. 

The fitting error on the parameterized $\sigma'$ from $\Delta V^{1/2}\lbrb{4}$ is treated as a systematic error, which is 70\%, since $\sigma'$ is consistent with zero in less than 2-$\sigma$ standard deviation. Similarly, the fitting errors on the parameters used in the $\Delta\eta$-dependent correlation $D$ are treated as systematic errors that are propagated through to the total uncertainty on $D$. In addition, there is a systematic error on $D$ that is associated with the choice of fitting function shown as Eq.~\eqref{Eqd}, the magnitude of which was estimated using different forms of the fitting function. The forms tried included: an exponential term plus a linear term, a Gaussian function plus a linear term, an exponential function only, a Gaussian function only, and an exponential function plus a term of the form $e^{-\frac{1}{2}(\frac{\Delta\eta}{\sigma})^{4}}$. 

The total estimated uncertainty in the second harmonic of $D(\Delta\eta)$ is an average of 40\% based on the different sources evaluated. This systematic error on $D$ also applies to the decomposed flow through $\mn{v^{2}} = V\lbrb{2} - D$.

\section{Results and Discussion}
\label{sec:result}

\begin{figure*}[htb]	
 \centering
  \includegraphics[width=0.4\textwidth]{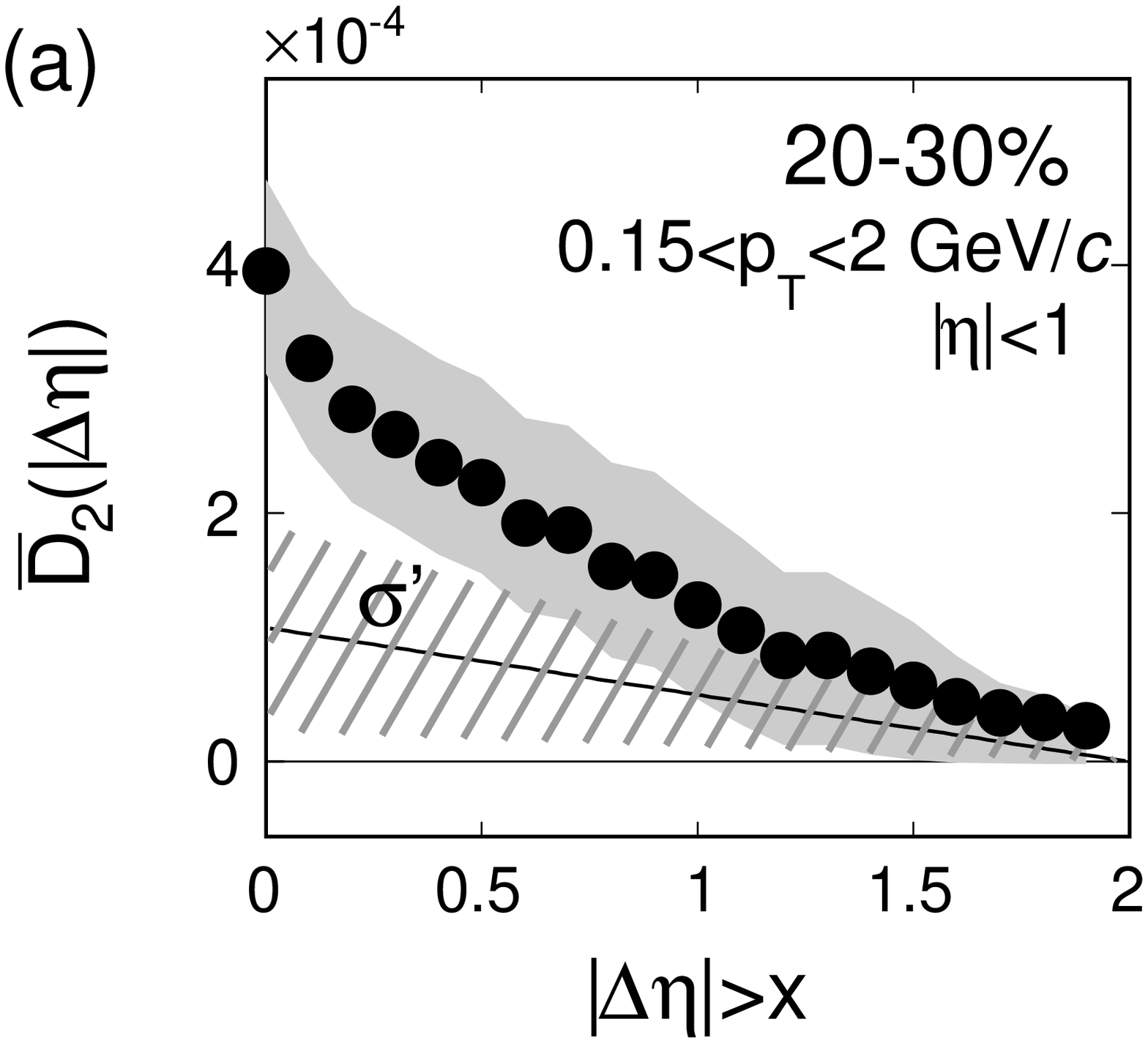} 
  \includegraphics[width=0.4\textwidth]{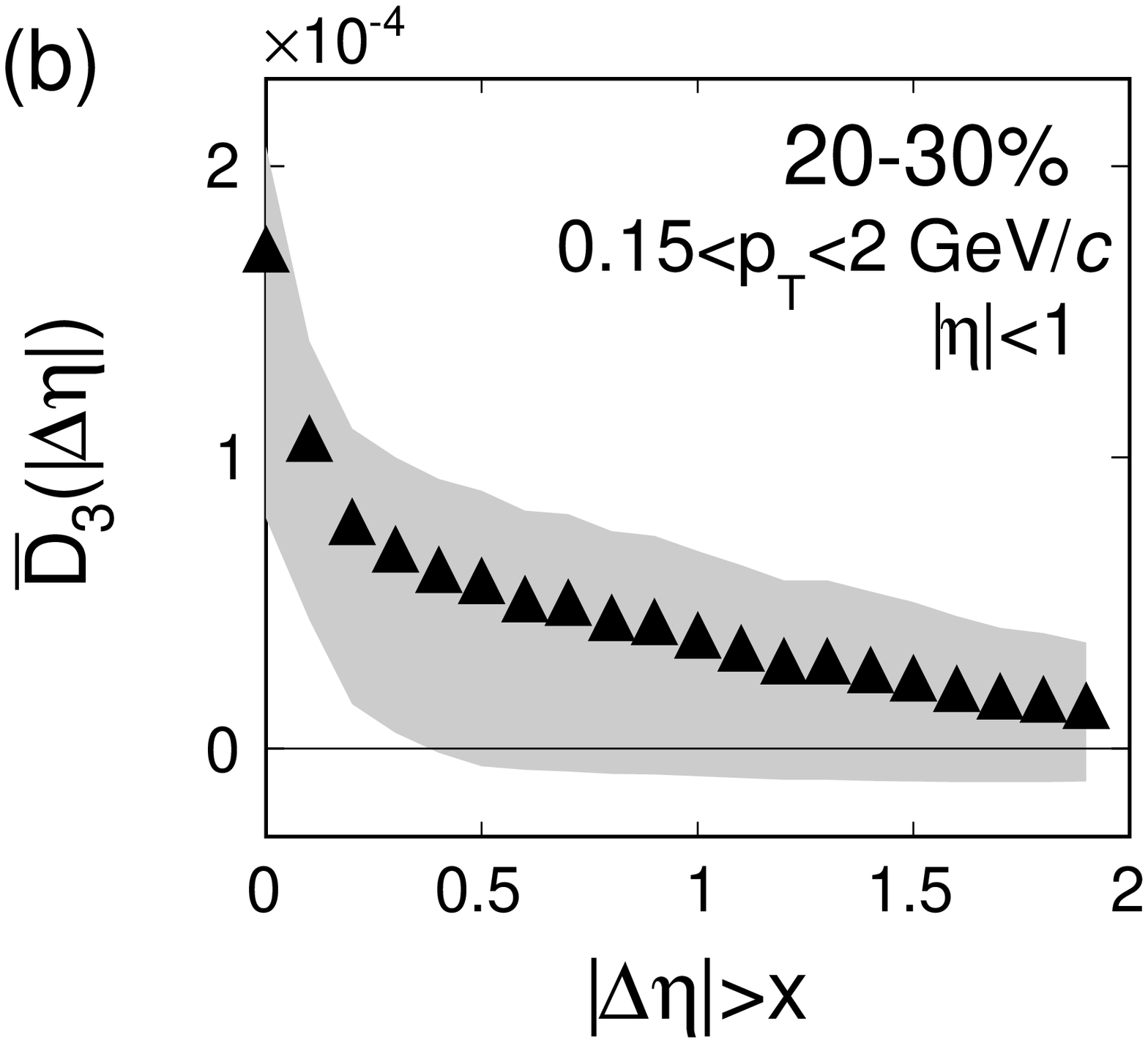} 
  
 \caption{The $\Delta\eta$-dependent component of the two-particle cumulant with $\Delta\eta$-gap, $\bar{D}$ in Eq.~\eqref{Eqaved}, of the second (a) and third (b) harmonics is shown as a function of $\Delta\eta$-gap $|\Delta\eta| > x$. ($x$ is the x-axis value.) The shaded bands are systematic uncertainties. In (a) the estimated $\sigma'$ is indicated as the straight line, with its uncertainty of $\pm$1 standard deviation as the cross-hatched area for 20-30\% central Au+Au collisions at $\sqrt{s_{_{\rm NN}}} =$ 200 GeV.  }
  \label{figetagap}
\end{figure*} 

\begin{figure*}[htb]	
 \centering
 \begin{tabular}{cc}
  \includegraphics[width=0.4\textwidth]{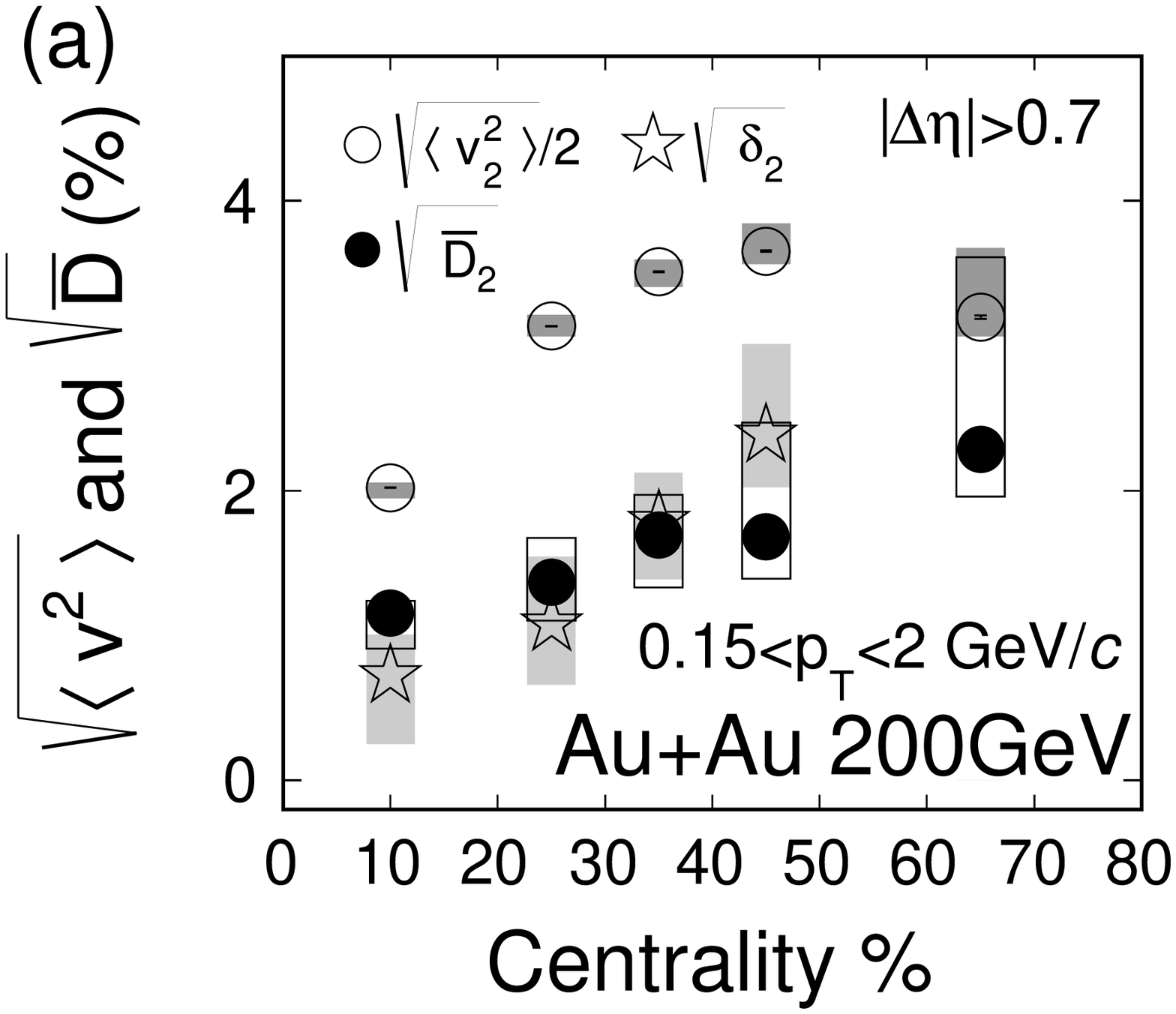} 
  \includegraphics[width=0.4\textwidth]{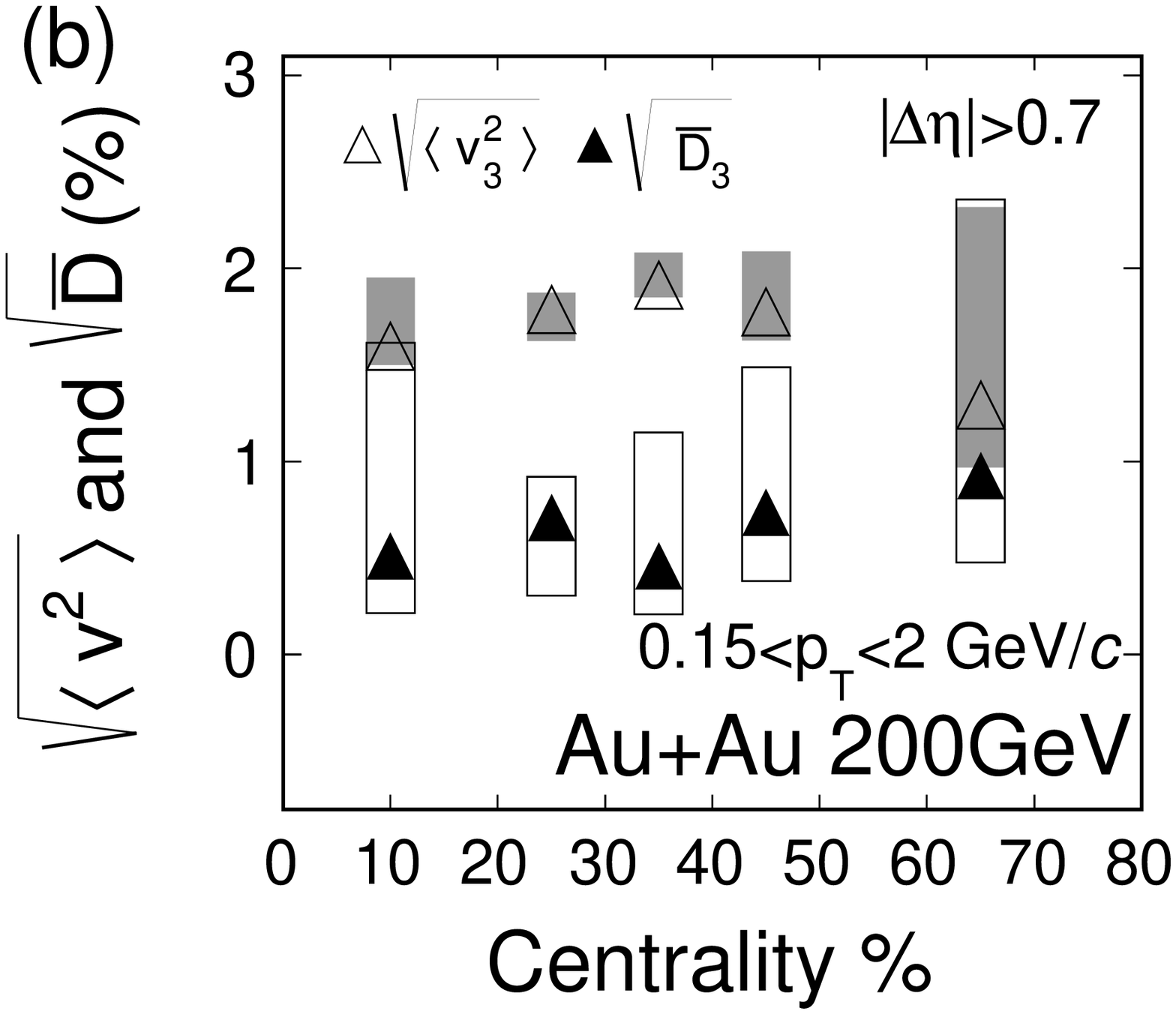} 
 \end{tabular}
 \caption{The nonflow, $\sqrt{\bar{D_{2}}}$ (solid dots), $\sqrt{\delta_{2}}$ (open stars), $\sqrt{\bar{D_{3}}}$ (solid triangles) and flow, $\sqrt{\mn{v_{2}^{2}}}/2$ (open circles), $\sqrt{\mn{v_{3}^{2}}}$ (open triangles) results are shown as a function of centrality percentile for the second (a) and third (b) harmonics, respectively. The statistical errors are smaller than the symbol sizes. The systematic errors are denoted by the vertical rectangles. }
\label{figetagapvscent}
\end{figure*}


Fig.~\ref{figv}(a) and (b) show the decomposed flow with flow fluctuations $\mn{v(\eta_{\alpha}) v(\eta_{\beta})}$ (see Eq.~\eqref{EqV2}) for $v_{2}$ and $v_{3}$, respectively. The results are found to be independent of $\eta$ for the measured pseudorapidity range $|\eta|<1$. The observed decrease of $V\lbrace2\rbrace$ in Fig.~\ref{figlegoV2V4} with increasing $\Delta\eta$ off diagonal is due to contributions from nonflow and $\Delta\eta$-dependent fluctuations. Note that the analysis method does not make any assumption about the $\eta$ dependence of flow; the flow can be $\Delta\eta$-independent but $\eta$-dependent. The observation that the decomposed flow and flow fluctuations are independent of $\eta$ is, therefore, significant.

Fig.~\ref{figv}(c) and (d) shows the projections of $\mn{v(\eta_{\alpha}) v(\eta_{\beta}))}$ in Fig.~\ref{figv}(a) and (b) onto one $\eta$ variable. The shaded band shows the systematic uncertainty, dominated by the systematic errors in the subtracted $D(\Delta\eta)$ term. For comparison, the projection of the $V_{2}\lbrace2\rbrace$ is also shown, where the shaded band is the systematic uncertainty. The projections are the respective quantities as a function of $\eta$ of one particle averaged over all $\eta$ of the other particle. The flows with $\Delta\eta$-independent fluctuation averaged over $\eta$ are $\sqrt{\mn{v_{2}^{2}}} = 6.27 \%\pm 0.003 \%(stat.) ^{+ 0.08}_{-0.07} \%(sys.) $ and $\sqrt{\mn{v_{3}^{2}}} = 1.78 \%\pm 0.008 \%(stat.) ^{+ 0.09}_{-0.16} \%(sys.) $ for our $p_{T}$ range $0.15 < p_{T} < 2 $ GeV/$c$ in the 20-30\% collision centrality range. The quoted statistical errors are from the $V\lbrb{2}$ measurements, while the systematic errors are dominated by the parameterization of $D$. The difference between  $V\lbrb{2}$ and $\mn{v^{2}}$ in Fig.~\ref{figv}(c) is the $D(\eta)$ versus $\eta$ of one particle averaged over all $\eta$ of the other particle. 

Figure \ref{figv}(c) also shows  the $V_{2}^{1/2}\lbrace4\rbrace$ projection as a function of $\eta$ as the solid blue triangles. $V_{2}^{1/2}\lbrace4\rbrace$ is also independent of $\eta$. The cyan band shows $V_{2}^{1/2}\lbrace4\rbrace +\sigma' = \mn{v}^{2}-\sigma^{2}$, with the systematic uncertainty that is dominated by the fitting uncertainty in $\sigma'$. The difference between the decomposed $\mn{v^{2}} = \mn{v}^{2} + \sigma^{2}$ and $V_{2}^{1/2}\lbrb{4}+\sigma'=\mn{v}^{2}-\sigma^{2}$ is the flow fluctuation, which is also independent of $\eta$ within the measured acceptance. The relative elliptic flow fluctuation is given by

\bea
  \dfrac{\sigma_{2}}{\mn{v_{2}}} & = & \sqrt{\dfrac{\mn{v_{2}^{2}}-(V_{2}^{\frac{1}{2}}\lbrb{4}+\sigma')}{\mn{v_{2}^{2}}+(V_{2}^{\frac{1}{2}}\lbrb{4} + \sigma')}} \nonumber \\
	  & = &34\% \pm 2\% (stat.) \pm 3\% (sys.),
\eea
where the systematic error is dominated by those in the parameterization of $D$ and $\sigma'$. The measured relative fluctuation is consistent with that from the PHOBOS experiment \cite{PHOBOS_sigma} and the previous STAR upper limit measurement \cite{STAR_sigma}.

Often, a $\Delta\eta$-gap is applied to reduce nonflow contamination in flow measurements. The nonflow $\bar{D}(|\Delta\eta|)$ with the $\Delta\eta$-gap is calculated as:
\bea
	\bar{D}(|\Delta\eta|) = \dfrac{ \int_{|\Delta\eta|}^{2} \mathrm{d} \Delta\eta' D(\Delta\eta')}{2-|\Delta\eta|}.
\label {Eqaved}
\eea
$|\Delta\eta| = 2$ is the acceptance limit in this analysis. $\bar{D}$ is the average of $D$ with $|\Delta\eta|$ larger than a certain value. Figure \ref{figetagap} (a) and (b) shows  $\bar{D}(|\Delta\eta|)$ as a function of $\Delta\eta$-gap $|\Delta\eta|>x$ ($x$ is the x-axis value) for the second and third harmonics, respectively. 
The bands are the systematic errors estimated from the fitting errors and the different fitting functions as described previously. These errors are correlated because, for all the points shown, the errors are calculated from the same parameters in the function $D$.

As noted above, $\bar{D}(|\Delta\eta|)$ is comprised of two parts: the contribution from the $\Delta\eta$-dependent flow fluctuation, $\sigma'$, and the term representing the nonflow, $\delta$. In Fig.~\ref{figetagap}(a), these two contributors are estimated separately. The straight line is an estimate of $\sigma'$. The cross-hatched area is its uncertainty of $\pm$1 standard deviation. The difference between the black solid points $\bar{D}(|\Delta\eta|)$ and the straight line $\sigma'$ is the nonflow contribution. For both the second harmonic and the third harmonic shown in Fig.~\ref{figetagap}(a) and Fig.~\ref{figetagap}(b), respectively, $\bar{D}(|\Delta\eta|)$ decreases as the $\Delta\eta$-gap between two particles increases. When $|\Delta\eta|>0.6$, $\bar{D}(|\Delta\eta|)$ is reduced to half of its value when $|\Delta\eta|>0$.

Figure \ref{figetagapvscent} shows $\sqrt{\mn{v^{2}}}$ and $\sqrt{\bar{D}}$ for all measured centralities for the second harmonic (a) and the third harmonic (b). $|\Delta\eta| > 0.7$ \cite{STAR_eventplane} is used to present the $\bar{D}$ result. The errors on $\sqrt{\mn{v^{2}}}$ and $\sqrt{\bar{D}}$ are anti-correlated. Taking $|\Delta\eta| > 0.7$, the relative magnitude $\bar{D_{2}}/\mn{v_{2}^{2}} =  5\% \pm 0.004 \% (stat.) \pm 2 \%(sys.)$ for 20-30\% centrality. It is clear that $\bar{D_{2}}$ increases as the collisions become  more peripheral.

\begin{figure}[htb]	
 \centering
 \begin{tabular}{cc}
  \includegraphics[width=0.4\textwidth]{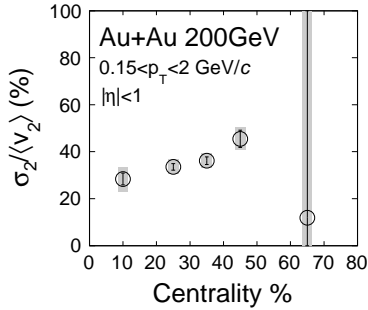} 
 \end{tabular}
 \caption{The relative elliptic flow fluctuation $\sigma_{2}/\mn{v_{2}}$ centrality dependence in $\sqrt{s_{_{\rm NN}}} =$ 200 GeV Au+Au collisions. The statistical errors are shown by the error bars. The systematic errors are denoted by the vertical rectangles. }
\label{figflucvscent}
\end{figure} 

The $\Delta\eta$-dependent nonflow contribution is mainly caused by near-side (small $\Delta\phi$) correlations. These correlations include jet-like correlations and resonance decays which decrease with increasing $\Delta\eta$.  The $\Delta\eta$-independent correlation is dominated by anisotropic flow. However, there should be a $\Delta\eta$-independent contribution from nonflow, such as away-side dijet correlations. This contribution should be smaller than the near-side nonflow contribution, because, in part, some of the away-side jets are outside the acceptance and, therefore, undetected.

Figure \ref{figflucvscent} shows $\sigma_{2}/\mn{v_{2}}$ for all measured centralities. From the central to the peripheral collisions, the relative elliptic flow fluctuation slightly increases. The statistics are limited in the most peripheral centrality bin.

\section{Summary}
\label{sec:summary}
We have analyzed two- and four-particle cumulant azimuthal anisotropies between pseudorapidity bins in Au+Au collisions at $\sqrt{s_{\rm NN}}=$200 GeV from STAR.  Exploiting the collision symmetry about midrapidity, we isolate the $\Delta\eta$-dependent and the $\Delta\eta$-independent azimuthal correlations in the data. The $\Delta\eta$-independent correlation, $\mn{v^{2}}$, is dominated by flow and flow fluctuations, and is found to be $\eta$ independent within the measured range of $\pm$1 unit of pseudorapidity. In the 20-30\% centrality Au+Au collisions, the elliptic flow fluctuation is found to be $\sigma_{2}/\mn{v_{2}} = 34\% \pm 2\% (stat.) \pm 3\% (sys.)$. The $\Delta\eta$-dependent correlation, $D(\Delta\eta)$, which may be attributed to nonflow, is found to be  $\bar{D}_{2}/\mn{v_{2}^{2}} = 5\% \pm 2\% (sys.)$ at $|\Delta\eta| > 0.7$ for $0.15< p_{T} < 2$ GeV/$c$.

\newpage                

\section{Acknowledgements}

We thank the RHIC Operations Group and RCF at BNL, the NERSC Center at LBNL, the KISTI Center in Korea, and the Open Science Grid consortium for providing resources and support. This work was supported in part by the Offices of NP and HEP within the U.S. DOE Office of Science, the U.S. NSF, CNRS/IN2P3, FAPESP CNPq of Brazil,  the Ministry of Education and Science of the Russian Federation, NNSFC, CAS, MoST and MoE of China, the Korean Research Foundation, GA and MSMT of the Czech Republic, FIAS of Germany, DAE, DST, and CSIR of India, the National Science Centre of Poland, National Research Foundation (NRF-2012004024), the Ministry of Science, Education and Sports of the Republic of Croatia, and RosAtom of Russia.












\end{document}